\documentstyle[12pt,epsfig]{article}
\def\be{\begin{equation}}
\def\ee{\end{equation}}
\def\bea{\begin{eqnarray}}
\def\eea{\end{eqnarray}}

\def\bar{\overline}

\def\l{\lambda}
\def\bc{\begin{center}}
\def\ec{\end{center}}
\def\O{{\cal O}}

\def\PR#1#2#3{Phys. Rev.  {\bf #1}, (#3) #2}
\def\PRL#1#2#3{Phys. Rev. Lett. {\bf #1}, (#3) #2}
\def\PL#1#2#3{Phys. Lett. {\bf #1}, (#3) #2}

\def\NP#1#2#3{Nucl. Phys. {\bf #1}, (#3) #2}

\def\PTP#1#2#3{Prog. Theor. Phys. {\bf #1}, (#3) #2}

\begin{document} 

\begin{flushright}
hep-ph/0209004 \qquad AUE-02-02 / KGKU-02-02 
\end{flushright}

\vspace{3mm}

\begin{center}
{\large \bf Precocious Gauge Symmetry Breaking \\
          in $SU(6) \times SU(2)_R$ Model }

\vspace{10mm}

Takemi HAYASHI 
            \footnote{E-mail address: hayashi@kogakkan-u.ac.jp}, 
Masato ITO$^a$ 
            \footnote{E-mail address: mito@eken.phys.nagoya-u.ac.jp}, \\
Masahisa MATSUDA$^b$
            \footnote{E-mail address: mmatsuda@auecc.aichi-edu.ac.jp}
and Takeo MATSUOKA$^c$ 
            \footnote{E-mail address: matsuoka@kogakkan-u.ac.jp}

\end{center}

\vspace{10mm}

\begin{center}
{\it 
Kogakkan University, Ise, JAPAN 516-8555 \\
{}$^a$Department of Physics, Nagoya University, Nagoya, 
JAPAN 464-8602 \\
{}$^b$Department of Physics and Astronomy, Aichi University 
of Education, Kariya, Aichi, JAPAN 448-8542 \\
{}$^c$Kogakkan University, Nabari, JAPAN 518-0498 
}
\end{center}

\vspace{3mm}

\begin{abstract}
In the $SU(6) \times SU(2)_R$ string-inspired model, 
we evolve the couplings and the masses down from the string 
scale $M_S$ using the renormalization group equations and 
minimize the effective potential. 
This model has  the flavor symmetry including the binary 
dihedral group $\tilde{D}_4$. 
We show that the scalar mass squared of the gauge non-singlet 
matter field possibly goes negative slightly below the string scale. 
As a consequence, the  precocious radiative breaking of the gauge 
symmetry down to the standard model gauge group can occur. 
In the present model, the large Yukawa coupling which plays 
an important role in the symmetry breaking is identical with 
the colored Higgs coupling related to the longevity of the proton. 
\end{abstract}

\newpage 
%%%%%%  SECTION  1  %%%%%%%%%%%%%%%%%%%%%%%%%%%%%%%%%%%%%%%%
\section{Introduction}
In the minimal supersymmetric standard model(MSSM), 
it is well-known that the spontaneous breaking of the gauge 
symmetry $SU(2)_L \times U(1)_Y \rightarrow U(1)_{em}$ is 
caused around the electroweak scale by the radiative effect 
due to the large top Yukawa coupling.\cite{Inoue} 
On the other hand, in many supersymmetric GUT models,  it is 
assumed that by taking the wine-bottle type of the Higgs 
potential by hand, the spontaneous breaking of a large gauge 
symmetry such as $SU(5)$ or $SO(10)$ takes place via Higgs 
mechanism at high energies around $\O(10^{16}{\rm GeV})$. 
In order to clarify whether or not the spontaneous breaking of 
the large gauge symmetry occurs at such a large energy scale, 
we need to address the underlying string theory which yields 
the GUT-type models. 
The radiative breaking of the large gauge symmetry occurs 
if the mass squared of a gauge non-singlet scalar field 
goes negative precociously as one evolves down 
from the string scale. 
Then it is of importance to study whether or not the radiative 
effect due to the large Yukawa couplings resulting from 
the underlying theory causes the scalar mass squared to be 
driven negative at a large energy scale. 
In the extra $U(1)^2$ string-inspired model it has 
been already found that the radiative effect due to 
the large Yukawa couplings possibly breaks down one of the extra 
$U(1)$ gauge symmetries around $\O(10^{15}{\rm GeV})$.\cite{Zoglin}

In this paper we consider the $SU(6) \times SU(2)_R$ 
string-inspired model, which contains many phenomenologically 
attractive features.\cite{Matsu1,Matsu2,Matsu3,CKM,MNS,Fuzz,Anom} 
In this model we evolve couplings and masses down from 
the string scale $M_S$ using the renormalization group(RG) 
equations and minimize the effective potential. 
The purpose of this paper is to explore whether the gauge 
symmetry breaking occurs or not at very large energy scale. 
Studying the RG evolution from the string scale $M_S$, 
we show that the scalar mass squared of the gauge non-singlet 
matter field possibly goes negative slightly below the string 
scale. 
This implies that the precocious breaking of the gauge symmetry 
$SU(6) \times SU(2)_R$ can occur due to the radiative effect. 
In this model the large Yukawa coupling which plays an 
important role in the symmetry breaking is identical with 
the colored Higgs coupling related to the longevity of 
the proton. 
This symmetry breaking triggers off the subsequent symmetry 
breaking.\cite{Scale} 
Thus we obtain the sequential symmetry breaking 
\[
  SU(6) \times SU(2)_R \longrightarrow 
   SU(4)_{PS} \times SU(2)_L \times SU(2)_R \longrightarrow 
   G_{SM},
\]
where $SU(4)_{PS}$ and $G_{SM}$ represent 
the Pati-Salam $SU(4)$\cite{Pati} and the standard model 
gauge group, respectively.

In the framework of the string theory we are prohibited 
from adding extra matter fields by hand. 
In the effective theory from string,  the matter contents and 
the Lagrangian are strongly constrained due to the topological 
and the symmetrical structure of the compact space. 
This situation is in sharp contrast to the conventional 
GUT-type models. 
For instance, in the perturbative heterotic string we have 
no adjoint or higher representation matter(Higgs) fields. 
Also, in the context of the brane picture,  matter fields 
belong to the bi-fundamental or the anti-symmetric representations 
under the gauge group such as $SU(M) \times SU(N)$. 
In the present model, under $SU(6) \times SU(2)_R$, 
gauge non-singlet matter fields consist 
of $({\bf 15},{\bf 1})$, $({\bf 6^*},{\bf 2})$ and 
their conjugates. 
Within the rigid framework we have to find out the path 
from the string scale physics to the low-energy physics. 
>From this point of view we study the RG evolution of 
couplings and masses from the string scale and explore 
the hierarchical path of the gauge symmetry breaking.

To the $SU(6) \times SU(2)_R$ string-inspired model 
we introduce the flavor symmetry 
${\bf Z}_M \times {\bf Z}_N \times \tilde{D}_4$.\cite{Fuzz} 
The cyclic group ${\bf Z}_M$ and the binary dihedral group 
$\tilde{D}_4$ have R symmetries, while ${\bf Z}_N$ has not. 
Introduction of the binary dihedral group $\tilde{D}_4$ is 
motivated by the phenomenological observation that 
the R-handed Majorana neutrino mass for the third generation 
has nearly the geometrically averaged magnitude of $M_S$ and $M_Z$. 
Further, the binary dihedral flavor symmetry $\tilde{D}_4$ is 
an extention of the R-parity. 
In Ref.\cite{Anom}, solving the anomaly-free conditions under 
many phenomenological constraints coming from the particle spectra, 
we found a large mixing angle(LMA)-MSW solution with 
$(M, \ N)=(19, \ 18)$, in which appropriate flavor charges 
are assigned to the matter fields. 
In Refs.\cite{Fuzz,Anom} we have assumed that the scalar mass 
squared of the gauge non-singlet field goes negative 
slightly below $M_S$. 
The results are in good agreement with the experimental 
observations about fermion masses and mixings and also about 
hierarchical energy scales including the GUT scale, 
the $\mu$ scale and the Majorana mass scale of the R-handed 
neutrinos. 
Then we carry out the present analysis of the RG evolution of 
the scalar masses squared on the basis of  the 
$SU(6) \times SU(2)_R$ model with the flavor symmetry 
${\bf Z}_{19} \times {\bf Z}_{18} \times \tilde{D}_4$.

This paper is organized as follows. 
In section 2, after explaining main features of 
the $SU(6) \times SU(2)_R$ string-inspired model with 
the flavor symmetry 
${\bf Z}_{19} \times {\bf Z}_{18} \times \tilde{D}_4$, 
we exhibit the superpotential. 
We point out that if the soft scalar mass squared is 
driven negative, the spontaneous breaking of the gauge 
symmetry $SU(6) \times SU(2)_R$ down to $G_{SM}$ 
occurs in two steps sequentially. 
In section 3 we study the RG evolutions of couplings and 
masses down from $M_S$. 
It is found that the scalar mass squared of the gauge 
non-singlet matter field possibly goes negative slightly 
below the string scale. 
The final section is devoted to summary and discussion.

\vspace{10mm}

%%%%%  SECTION  2  %%%%%%%%%%%%%%%%%%%%%%%%%%%%%%%%%%%%%%%%%
\section{$SU(6) \times SU(2)_R$ Model and the Scalarpotential}

The $SU(6) \times SU(2)_R$ string-inspired model considered here 
is studied in detail in 
Refs.\cite{Matsu1,Matsu2,Matsu3,CKM,MNS,Fuzz,Anom}. 
To begin with, we review the main features of the model. 
\begin{enumerate}
\item The gauge group $G = SU(6) \times SU(2)_R$ can be 
obtained from $E_6$ through the ${\bf Z}_2$ flux breaking 
on a multiply-connected manifolds $K$.\cite{Hoso,Flux1,Flux2} 
To be more specific, the nontrivial holonomy $U_d$ on $K$ is 
of the form 
\be
    U_d = \exp \left( \pi i I_{3R} \right), 
\ee
where $I_{3R}$ represents the third direction of the $SU(2)_R$. 
The symmetry breaking of $G$ down to $G_{\rm SM}$ can take 
place via the Higgs mechanism without matter fields of 
adjoint or higher representations. 
$SU(6) \times SU(2)_R$ is the largest one of 
such gauge groups.\cite{Matsu4} 

\item Matter consists of the chiral superfields of three families 
and the one vector-like multiplet, i.e., 
\be
  3 \times {\bf 27}(\Phi_{1,2,3}) + 
        ({\bf 27}(\Phi_0)+\overline{\bf 27}({\bar \Phi})) 
\ee
in terms of $E_6$. 
The superfields $\Phi$ in {\bf 27} of $E_6$ are decomposed into 
the irreducible representations of $G = SU(6) \times SU(2)_R$ as 
\be
  \Phi({\bf 27})=\left\{
       \begin{array}{lll}
         \phi({\bf 15},{\bf 1})& : 
               & \quad \mbox{$Q,L,g,g^c,S$}, \\
          \psi({\bf 6}^*,{\bf 2}) & : 
               & \quad \mbox{$(U^c,D^c),(N^c,E^c),(H_u,H_d)$}, 
       \end{array}
       \right.
\label{eqn:27}
\ee
where $g$ and $g^c$ and $H_u$ and $H_d$ represent the colored 
Higgs and the doublet Higgs superfields, respectively, 
$N^c$ is the right-handed neutrino superfield, and 
$S$ is an $SO(10)$ singlet. 
It should be noted that the doublet Higgs and the color-triplet 
Higgs fields belong to the different irreducible representations 
of $G$ as shown in Eq.(\ref{eqn:27}). 
As a consequence, the triplet-doublet splitting problem is 
solved naturally.\cite{Matsu1} 

\item As the flavor symmetry, we introduce 
the ${\bf Z}_{19} \times {\bf Z}_{18}$ and the $\tilde{D}_4$ symmetries 
and regard ${\bf Z}_{19}$ and ${\bf Z}_{18}$ as the R and the non-R 
symmetries, respectively. 
Since the numbers 19 and 18 are relatively prime, 
we can combine these symmetries as 
\be
  {\bf Z}_{19} \times {\bf Z}_{18} = {\bf Z}_{342}. 
\ee
Solving the anomaly-free conditions under many 
phenomenological constraints coming from the particle spectra, 
we obtain a LMA-MSW solution with the ${\bf Z}_{342}$ charges 
of matter superfields as shown in Table 1.\cite{Anom} 
In this solution we assign the Grassmann number $\theta$, 
which has the charge $(-1, \ 0)$ under 
${\bf Z}_{19} \times {\bf Z}_{18}$, 
the charge 18 under ${\bf Z}_{342}$. 
The assignment of "$\tilde{D}_4$ charges" to matter superfields 
is given in Table 2, where $\sigma_i \, (i=1,2,3)$ represent 
the Pauli matrices and 
\be
   \sigma_4 =  \left(
       \begin{array}{cc}
         1  &  0  \\
         0  &  i  
       \end{array}
       \right). 
\ee 
The $\sigma_3$ transformation yields the R-parity. 
Namely, the R-parities of the superfields $\Phi_i \, (i=1,2,3)$ 
for three generations are all odd, 
while those of the $\Phi_0$ and $\bar{\Phi}$ are even.

%%%%%  TABLE  1  %%%%%%%%%%%%%%%%%%%%%%%%%%%%%%%
\begin{table}
\caption{Assignment of ${\bf Z}_{342}$ charges 
          for matter superfields}
\label{table:1}
\bc
\begin{tabular}{|c|ccc|cc|} \hline \hline 
  & \phantom{M} $\Phi_1$ \phantom{MM} & 
      \phantom{MM} $\Phi_2$ \phantom{MM} & 
        \phantom{MM} $\Phi_3$ \phantom{M} & 
          \phantom{M} $\Phi_0$ \phantom{MM} & 
            \phantom{MM} $\bar{\Phi}$ \phantom{M} \\ \hline
$\phi({\bf 15, \ 1})$   &  $a_1=126$  &  $a_2=102$  &  $a_3=46$  
                                &  $a_0=12$  &  $\bar{a}=-16$  \\
$\psi({\bf 6^*, \ 2})$  &  $b_1=120$  &  $b_2=80$  &  $b_3=16$  
                         &  $b_0=-14$  &  $\bar{b}=-67$  \\ \hline
\end{tabular}
\ec
\end{table}
%%%%%%%%%%%%%%%%%%%%%%%%%%%%%%%%%%%%%%%%%%%%%%%

%%%%%  TABLE  2  %%%%%%%%%%%%%%%%%%%%%%%%%%%%%%%
\begin{table}
\caption{Assignment of "$\tilde{D}_4$ charges" to matter superfields}
\label{table:2}
\bc
\begin{tabular}{|c|ccc|} \hline \hline 
    & \phantom{M} $\Phi_i \ (i=1,2,3)\ $ \phantom{M} & 
        \phantom{M} $\Phi_0$ \phantom{M} & 
              \phantom{M} $\bar{\Phi}$ \phantom{M} \\ \hline
$\phi({\bf 15, \ 1})$   & 
                   $\sigma_1$  &     1       &       1       \\
$\psi({\bf 6^*, \ 2})$  & 
                   $\sigma_2$  & $\sigma_3$  &   $\sigma_4$  \\  \hline
\end{tabular}
\ec
\end{table}
%%%%%%%%%%%%%%%%%%%%%%%%%%%%%%%%%%%%%%%%%%%%%%%%

\item There are two types of gauge invariant trilinear 
combinations 
\bea
    (\phi ({\bf 15},{\bf 1}))^3 & = & QQg + Qg^cL + g^cgS, \\
    \phi ({\bf 15},{\bf 1})(\psi ({\bf 6}^*,{\bf 2}))^2 & 
            = & QH_dD^c + QH_uU^c + LH_dE^c  + LH_uN^c 
                                            \nonumber \\ 
             {}& & \qquad   + SH_uH_d + 
                     gN^cD^c + gE^cU^c + g^cU^cD^c. 
\label{eqn:trico}
\eea
\end{enumerate}

The flavor symmetry requires that in the superpotential 
these trilinear combinations are multiplied by some powers of 
$\phi_0 \bar{\phi}$ or $\psi_0 \bar{\psi}$. 
Concretely, the superpotential terms are of the forms 
\bea
  W_Y & = & \frac{1}{3!} \, z_0 \left( \frac{\phi_0 \bar{\phi}}
              {M_S^2} \right)^{\zeta_0} (\phi_0)^3 
          + \frac{1}{3!} \, \bar{z} \left( \frac{\phi_0 \bar{\phi}}
              {M_S^2} \right)^{\bar{\zeta}} (\bar{\phi})^3 
          + \frac{1}{2} \, h_0 \left( \frac{\phi_0 \bar{\phi}}{M_S^2} 
               \right)^{\eta_0} \phi_0 \psi_0 \psi_0   \nonumber \\
     {} & & \phantom{M} + \frac{1}{2} \bar{h} \left( 
               \frac{\phi_0 \bar{\phi}}{M_S^2} \right)^{\bar{\eta}} 
                  \left( \frac{\psi_0 \bar{\psi}}{M_S^2} \right)^2 
                    \bar{\phi} \, \bar{\psi} \, \bar{\psi} 
           + \sum_{i,j=1}^{3} z_{ij} \left( \frac{\phi_0 \bar{\phi}}
           {M_S^2} \right)^{\zeta_{ij}} \phi_0 \phi_i \phi_j   \nonumber \\
     {}& & \phantom{M} + \sum_{i,j=1}^{3} h_{ij} \left( 
           \frac{\phi_0 \bar{\phi}}{M_S^2} \right)^{\eta_{ij}} 
                          \phi_0 \psi_i \psi_j 
     + \sum_{i,j=1}^{3} m_{ij} \left( \frac{\phi_0 \bar{\phi}}
              {M_S^2} \right)^{\mu_{ij}} \psi_0 \phi_i \psi_j. 
\label{eqn:WY}
\eea
The exponents are determined by the constraints coming from 
the flavor symmetry. 
To be more specific, we have 
\bea
    (\zeta_0, \ \bar{\zeta}, \ \eta_0, \ \bar{\eta}) 
              = (0, \ 150, \ 158, \ 84), \qquad 
    \zeta_{ij} =  \left(
       \begin{array}{ccc}
         57  &  51  &  37  \\
         51  &  45  &  31  \\
         37  &  31  &  17  
       \end{array}
       \right)_{ij},       \nonumber \\
    \eta_{ij} =  \left(
       \begin{array}{ccc}
         54  &  44  &  28  \\
         44  &  34  &  18  \\
         28  &  18  &   2  
       \end{array}
       \right)_{ij},        \phantom{MMMM} 
    \mu_{ij} =  \left(
       \begin{array}{ccc}
         49  &  39  &  23  \\
         43  &  33  &  17  \\
         29  &  19  &   3  
       \end{array}
       \right)_{ij}. 
\eea
The coefficients $z_0$, $\bar{z}$, $h_0$, $\bar{h}$, $z_{ij}$, 
$h_{ij}$ and $m_{ij}$ are $\O(1)$ constants. 
The relation $\zeta_0 = 0$ means that only the superfield 
$\phi_0$ takes part in the renormalizable interaction with the 
large Yukawa coupling at $M_S$. 
The powers of $\phi_0 \bar{\phi}$ can be replaced all or in part 
with those of $\psi_0 \bar{\psi}$ subject to the flavor symmetry. 
When $\phi_0$ and $\bar{\phi}$ develop the non-zero 
vacuum expectation values(VEV's), 
the above non-renormalizable terms become the effective Yukawa 
couplings with hierarchical patterns.\cite{F-N} 
In addition, the superpotential contains the other types of 
the non-renormalizable term
\be
  W_1 = M_S^3 \left[ \l_0 
     \left( \frac{\phi_0 \bar{\phi}}{M_S^2} \right)^{2n} 
       + \l_1 \left( \frac{\phi_0 \bar{\phi}}{M_S^2} \right)^n 
         \left( \frac{\psi_0 \bar{\psi}}{M_S^2} \right)^m 
       + \l_2 \left( \frac{\psi_0 \bar{\psi}}{M_S^2} \right)^{2m}\right] 
\label{eqn:W1}
\ee
with $\l_i = \O(1)$. 
The flavor symmetry ${\bf Z}_{342} \times \tilde{D}_4$ yields 
$n = 81$ and $m=4$.

In the present model we study the minimum point of 
the scalarpotential. 
We will assume that the supersymmetry is broken at the string 
scale due to the hidden sector dynamics and that 
the supersymmetry breaking is communicated gravitationally 
to the observable sector via the soft supersymmetry breaking terms. 
As mentioned above, there exists a large Yukawa coupling 
at the string scale only for $\phi_0$. 
Then the radiative corrections of the soft scalar masses squared 
due to the Yukawa coupling are sizable only for $\phi_0$. 
On the other hand, the R-parity odd superfields $\phi_i$ and $\psi_i$ 
$(i=1,2,3)$ which appear in pair in the superpotential term, 
hardly receive the radiative corrections in the region around $M_S$. 
The soft scalar masses squared of the R-parity odd superfields 
remain positive in the wide energy region. 
Therefore, the F-flat conditions for $\phi_i$ and $\psi_i$ 
$(i=1,2,3)$ require 
\be
   \langle \phi_i \rangle = \langle \psi_i \rangle = 0. 
         \phantom{MMM} (i=1,2,3)
\ee
Thus in order to minimize the scalarpotential, 
it is sufficient for us to confine ourselves to the R-parity 
even sector. 
In the R-parity even sector we have the superpotential $W_1$. 
The scalarpotential is given by 
\bea
  V & = & \left| \frac{\partial W_1}{\partial \phi_0} \right|^2 
      + \left| \frac{\partial W_1}{\partial \bar{\phi}} \right|^2 
      + \left| \frac{\partial W_1}{\partial \psi_0} \right|^2 
      + \left| \frac{\partial W_1}{\partial \bar{\psi}} \right|^2 
                        \nonumber  \\
      & & \phantom{MMMMM} 
        + ({\rm D \ term}) + V_{\rm soft} + \Delta V_{\rm 1-loop} \, ,  
\label{eqn:Pot}         
\eea
where $V_{\rm soft}$ represents the soft supersymmetry breaking 
terms 
\be
   V_{\rm soft} = \tilde{m}_{\phi 0}^2 \, | \phi_0 |^2 
                   + \tilde{m}_{\bar{\phi}}^2 \, | \bar{\phi} |^2 
                   + \tilde{m}_{\psi 0}^2 \, | \psi_0 |^2 
                   + \tilde{m}_{\bar{\psi}}^2 \, | \bar{\psi} |^2 
                   + ({\rm A \ term}). 
\label{eqn:Vsoft}
\ee
The one-loop correction $\Delta V_{\rm 1-loop}$ is of 
the form\cite{CW} 
\be
   \Delta V_{\rm 1-loop} = 
                   \frac{1}{64 \pi^2} {\rm STr} \, {\cal M}^4 
            \left[ \ln \left( \frac{{\cal M}^2}{Q^2} \right) 
                  - \frac{3}{2} \right], 
\label{eqn:1-loop}
\ee
where bosons(fermions) contribute with positive(negative) sign in 
the supertrace and the mass ${\cal M}$ has to be considered to be 
a function of $\phi_0$, $\bar{\phi}$, $\psi_0$ and $\bar{\psi}$. 
In the above equations, for simplicity we denote the scalar 
components of the superfields by the same letters as 
the superfields. 
The soft scalar masses squared $\tilde{m}_{\phi 0}^2$, 
$\tilde{m}_{\bar{\phi}}^2$, $\tilde{m}_{\psi 0}^2$ and 
$\tilde{m}_{\bar{\psi}}^2$ are assumed to take a universal 
positive value at the string scale. 
We evolve down from the string scale the scalar masses squared 
by using the RG equations. 
Since the one-loop correction Eq.(\ref{eqn:1-loop}) is linear in 
$\ln Q^2$, it is possible to find a value of $Q$ such that 
this correction is quite small in the minimum of the potential. 
Therefore, it is sufficient to treat the minimization by simply 
using the RG-improved tree-level potential. 
If $\tilde{m}_{\phi 0}^2 + \tilde{m}_{\bar{\phi}}^2$ is driven 
negative slightly below the string scale, 
the gauge symmtry could be spontaneously broken irrespective 
of $\tilde{m}_{\psi 0}^2$ and $\tilde{m}_{\bar{\psi}}^2$. 
By minimizing the scalarpotential in the case 
$\tilde{m}_{\phi 0}^2 + \tilde{m}_{\bar{\phi}}^2 < 0$, 
we can determine the energy scales of the gauge symmetry breaking, 
that is, the VEV's 
$\langle \phi_0 \rangle$, $\langle \bar{\phi} \rangle$, 
$\langle \psi_0 \rangle$ and $\langle \bar{\psi} \rangle$. 
The D-flat conditions require 
\be
  |\langle \phi_0 \rangle| = |\langle \bar{\phi} \rangle|, \qquad 
  |\langle \psi_0 \rangle| = |\langle \bar{\psi} \rangle|.
\ee
Thus, if $\tilde{m}_{\phi 0}^2 + \tilde{m}_{\bar{\phi}}^2 < 0$, 
the minimum point of the scalarpotential becomes\cite{Scale} 
\bea
   |\langle \phi_0 \rangle| = |\langle \bar{\phi} \rangle| 
    & \sim & M_S \, \left( \frac{\tilde{m}_{\phi}}{M_S} 
              n^{-3/2} \right)^{1/(4n-2)},  \\
   |\langle \psi_0 \rangle| = |\langle \bar{\psi} \rangle| 
    & \sim & M_S \, \left( \frac{|\langle \phi_0 \rangle|}{M_S} 
               \right)^{n/m}
\eea
in a feasible parameter region of the coefficients $\l_i$, 
where $\tilde{m}_{\phi} = 
\sqrt{|\tilde{m}_{\phi 0}^2 + \tilde{m}_{\bar{\phi}}^2|}$. 
Since we obtain 
$|\langle \phi_0 \rangle| > |\langle \psi_0 \rangle|$, 
the gauge symmetry breaking occurs in two steps as 
\be
  SU(6) \times SU(2)_R \longrightarrow 
   SU(4)_{PS} \times SU(2)_L \times SU(2)_R \longrightarrow 
   G_{SM}.
\ee
When the gauge symmetry $SU(6) \times SU(2)_R$ is broken down to 
$SU(4)_{PS} \times SU(2)_L \times SU(2)_R$, 
the field $\phi_0({\bf 15, \ 1})$ is decomposed as 
\be
   \phi_0({\bf 15, \ 1}) \longrightarrow 
   \phi_0({\bf 4, \ 2, \ 1}), \quad 
   \phi_0({\bf 6, \ 1, \ 1}), \quad 
   \phi_0({\bf 1, \ 1, \ 1}). 
\ee
Needless to say, the field $\phi_0({\bf 1, \ 1, \ 1})$ develops 
the non-zero VEV $|\langle \phi_0 \rangle|$. 
In addition, the field $\psi_0({\bf 6^*, \ 2})$ is decomposed as 
\be
   \psi_0({\bf 6^*, \ 2}) \longrightarrow 
   \psi_0({\bf 4^*, \ 1, \ 2}), \quad 
   \psi_0({\bf 1, \ 2, \ 2}). 
\ee
A question arises as to which field of 
$\psi_0({\bf 4^*, \ 1, \ 2})$ and $\psi_0({\bf 1, \ 2, \ 2})$ 
develops the non-zero VEV $|\langle \psi_0 \rangle|$. 
As seen in Eq.(\ref{eqn:trico}), 
the field $\psi_0({\bf 1, \ 2, \ 2})$ has the coupling 
$\phi_0({\bf 1, \ 1, \ 1}) \psi_0({\bf 1, \ 2, \ 2})^2$, 
which is the third term of Eq.(\ref{eqn:WY}). 
Below the scale $|\langle \phi_0 \rangle|$,  this term induces 
th $\mu$ term. 
The F-flat condition for $\psi_0({\bf 1, \ 2, \ 2})$ requires 
\be
  |\psi_0({\bf 1, \ 2, \ 2})| = 0
\ee
at a large energy scale. 
Consequently, the non-zero VEV $|\langle \psi_0 \rangle|$ is 
attributed to $\psi_0({\bf 4^*, \ 1, \ 2})$. 
Thus we obtain $G_{SM}$ below the scale $|\langle \psi_0 \rangle|$.

We now calculate the energy scale of the gauge symmetry 
breaking. 
Since we have $n=81$ and $m=4$, 
the VEV's $|\langle \phi_0 \rangle|$ and $|\langle \psi_0 \rangle|$ 
are smaller than $M_S$ but not far from $M_S$. 
By taking $M_S \sim 5 \times 10^{17}{\rm GeV}$\cite{MS} 
and $\tilde{m}_{\phi} \sim 10^3{\rm GeV}$, 
we obtain 
\be
   \frac{|\langle \phi_0 \rangle|}{M_S} \simeq 0.89, \qquad 
   \frac{|\langle \psi_0 \rangle|}{M_S} \simeq 0.10. 
\ee
Therefore, for the present model to be consistent, 
it is necessary that 
$\tilde{m}_{\phi 0}^2 + \tilde{m}_{\bar{\phi}}^2$ 
is deriven negative slightly below $M_S$. 
In the next section, evolving couplings and masses down 
from the string scale using the RG equations, 
we show that $\tilde{m}_{\phi 0}^2 + \tilde{m}_{\bar{\phi}}^2$ 
possibly goes negative slightly below the string scale.

\vspace{10mm}

%%%%%  SECTION  3  %%%%%%%%%%%%%%%%%%%%%%%%%%%%%%%%%%%%%%%%%
\section{The RG evolutions of scalar masses}
The one-loop RG equation for the $SU(6)$ gauge coupling 
$g_6$ is given by 
\be
  (4 \pi)^2 \frac{dg_6^2}{dt} = - 2 b_6 \, g_6^4, 
\ee
where the variable $t$ is defined as $\ln(Q/M_S)$ and $b_6 = 3$. 
Similarly, we have the one-loop RG equation for the $SU(6)$ 
gaugino mass $M_6$ 
\be
  (4 \pi)^2 \frac{dM_6}{dt} = - 2 b_6 \, g_6^2 \, M_6. 
\ee
These equations are easily solved as 
\be
   g_6^2(u) = \frac{g_0^2}{u}, \qquad 
   M_6(u)   = \frac{M_{1/2}}{u}, 
\ee
where the variable $u$ is defined as 
\be
   u = 1 + \frac{3}{8\pi^2} g_0^2 \, t. 
\ee
The constants $g_0$ and $M_{1/2}$ represent the values of 
the gauge coupling and the gaugino mass at the string scale 
$M_S$, respectively.

At the string scale the renormalizable term of 
the superpotential is of the simple form 
\be
   W_{Y0} = \frac{1}{3!} \, z_0 (\phi_0)^3. 
\label{eqn:WY0}
\ee
This term induces the soft breaking A term 
\be
   V_{\rm soft} \supset \frac{1}{3!} \, A_0 \, z_0 (\phi_0)^3.
\ee
Here we assume that the Yukawa coupling $z_0$ and the soft 
breaking parameter $A_0$ are real. 
In this case the RG equations for $z_0$,  $A_0$, $\tilde{m}_{\phi 0}^2$ 
and $\tilde{m}_{\bar{\phi}}^2$ are given by\cite{Falck}
\bea
  (4 \pi)^2 \frac{dz_0^2}{dt} & = & 
                  \left( -56 \, g_6^2 + 18 \, z_0^2 \right) z_0^2, 
\label{eqn:z02}                    \\
  (4 \pi)^2 \frac{dA_0}{dt} & = & 
                  3 A_0 \, z_0^2 + 56 M_6 \, g_6^2, 
\label{eqn:A0}                     \\
  (4 \pi)^2 \frac{d \tilde{m}_{\phi 0}^2}{dt} & = & 
     6 \left( \tilde{m}_{\phi 0}^2 + \frac{1}{3} A_0^2 \right) z_0^2 
         - \frac{112}{3} M_6^2 \, g_6^2, 
\label{eqn:mphi2}                 \\
  (4 \pi)^2 \frac{d \tilde{m}_{\bar{\phi}}^2}{dt} & = & 
                             - \frac{112}{3} M_6^2 \, g_6^2. 
\label{eqn:mpsi2}
\eea
Concretely, the RG evolutions are expressed as 
\bea
  \frac{z_0^2(u)}{z_0^2(1)} & = & 
                               \frac{1}{u \, D(u)}, \\
  \frac{A_0(u)}{A_0(1)}     & = & 
                     \frac{u^{25/18}}{D(u)^{1/6}} \, E(u), \\
  \frac{\tilde{m}_{\phi 0}^2(u)}{\tilde{m}_{\phi 0}^2(1)} & = & 
        \frac{u^{25/9}}{D(u)^{1/3}} \left[ 1 - r_1^2 F(u) \right], \\
  \frac{\tilde{m}_{\bar{\phi}}^2(u)}{\tilde{m}_{\bar{\phi}}^2(1)} & = & 
          1 + \frac{28}{9} \, r_1^2 \left( u^{-2} - 1 \right) 
\eea
with $\tilde{m}_{\phi 0}^2(1) = \tilde{m}_{\bar{\phi}}^2(1) 
\equiv \tilde{m}_0^2$. 
In these expressions we define three dimensionless parameters 
\be
   r_0 = \frac{z_0(1)}{g_0}, \qquad 
   r_1 = \frac{M_{1/2}}{\tilde{m}_0}, \qquad 
   r_2 = \frac{A_0(1)}{M_{1/2}}, 
\ee
and three functions 
\bea
   D(u) & = & \left( 1 - \frac{9}{25}r_0^2 \right) u^{25/3} 
                            + \frac{9}{25} r_0^2,  \\
   E(u) & = & 1 - \frac{28}{3 \, r_2} \, 
                  \int_u^1 \frac{ D(v)^{1/6}}{v^{61/18}} dv,  \\
   F(u) & = & \int_u^1 \left[ \frac{r_0^2 \, r_2^2}{3\, v \, D(v)} 
       \, E(v)^2 
           - \frac{56}{9} \, \frac{D(v)^{1/3}}{v^{52/9}} \right] dv. 
\eea

Since we are interested in the precocious breaking of the gauge 
symmetry, 
the RG evolution of the couplings and the masses is carried out 
in the energy region ranging from $M_S$ to $M_S/5$. 
Since $4\pi/g_0^2$ takes a value around 15 in the present 
model\cite{Matsu1}, we have $g_0 \simeq 0.9$ and then 
the region considered here of the variable $u$ becomes 
$1.0 \sim 0.95$. 
We now proceed to accomplish the numerical study as to whether 
or not $\tilde{m}_{\phi 0}^2 + \tilde{m}_{\bar{\phi}}^2$ 
is driven negative in the region $u = 1.0 \sim 0.95$. 
As a typical example, in Fig.1 we show the calculation of 
$(\tilde{m}_{\phi 0}^2(u) + \tilde{m}_{\bar{\phi}}^2(u))
/\tilde{m}_0^2$ 
for the parameter set 
$(r_0, \ r_1, \ r_2) = (3.0, \ 3.0, \ 3.0 \sim 4.0)$. 
We find that 
$\tilde{m}_{\phi 0}^2(u) + \tilde{m}_{\bar{\phi}}^2(u)$ 
is driven negative at $u \sim 0.98$ 
if the value of $r_2$ is larger than 3.3. 
In Fig.2 also 
$(\tilde{m}_{\phi 0}^2(u) + \tilde{m}_{\bar{\phi}}^2(u))
/\tilde{m}_0^2$ 
for the parameter set 
$(r_0, \ r_1, \ r_2) = (3.0, \ 2.5 \sim 4.0, \ 3.5)$ 
is given. 
>From these figures it turns out that the precocious breakdown 
is realized in the parameter region of the Yukawa coupling 
$z_0(1) \sim 2.7$ and $r_1, \ r_2 = 3.0 \sim 4.0$.

In the present choice of $z_0(1)$ we have 
$z_0(1)^2/(4 \pi) \sim  0.6$, which does not seem to be small 
enough to use it as the perturbative expansion parameter. 
However, $z_0(u)$ diminishes in magnitude with decreasing $u$. 
The present analysis is sufficient to show that 
in the feasible parameter region, 
$\tilde{m}_{\phi 0}^2(u) + \tilde{m}_{\bar{\phi}}^2(u)$ 
possibly goes negative slightly below $M_S$.

%%%%%%%%%%%%%%%%%%%%%%%%%%%%%%%%%%%%%%%%%%%%%%%%%
\vspace{10mm}
\begin{figure}[htb]
\begin{center}
\includegraphics{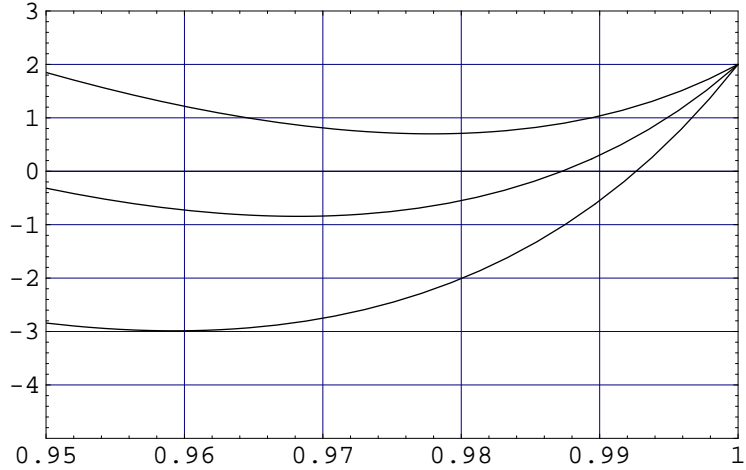}
\end{center}
\caption{
$(\tilde{m}_{\phi 0}^2(u) + 
\tilde{m}_{\bar{\phi}}^2(u))/\tilde{m}_0^2$ vs. $u$. 
The upper, middle and lower solid curves are for the cases of 
the parameter $r_2$ to be 3.0, 3.5, 4.0, respectively. 
Both the parameters $r_0$ and $r_1$ are taken as 3.0.
} 
\label{Fig1}
\end{figure}
%%%%%%%%%%%%%%%%%%%%%%%%%%%%%%%%%%%%%%%%%%%%%%%%%

%%%%%%%%%%%%%%%%%%%%%%%%%%%%%%%%%%%
\vspace{10mm}
\begin{figure}[htb]
\begin{center}
\includegraphics{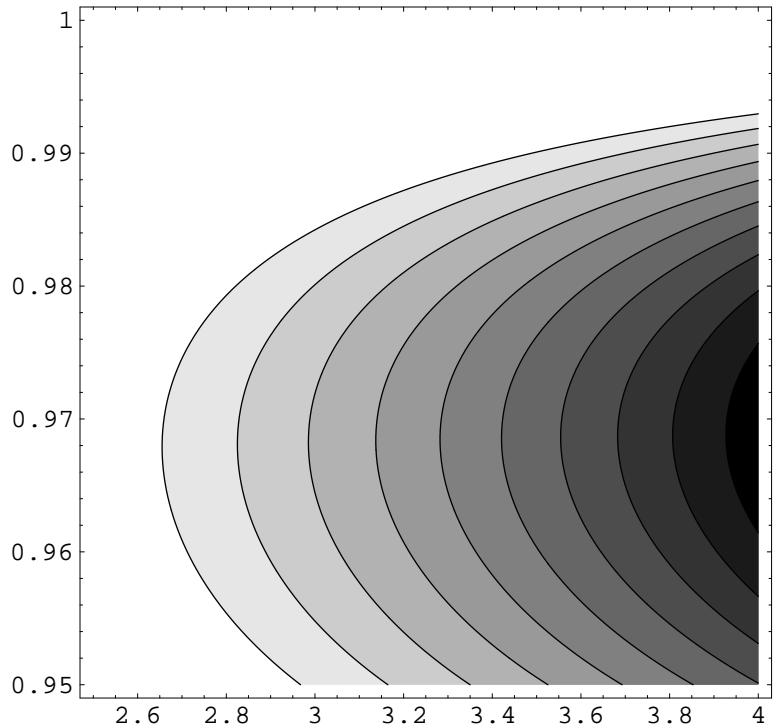}
\end{center}
\caption{
The $u$- and $r_1$-dependences of 
$(\tilde{m}_{\phi 0}^2(u) + \tilde{m}_{\bar{\phi}}^2(u))
/\tilde{m}_0^2$. 
The parameter $r_1$ varies from 2.5 to 4.0, 
while the parameters $r_0$ and $r_2$ are fixed as 3.0 and 3.5, 
respectively. 
In the white region  
$(\tilde{m}_{\phi 0}^2(u) + \tilde{m}_{\bar{\phi}}^2(u))
/\tilde{m}_0^2$ 
is positive and in the region from gray to black has 
a negative value up to $-3.0$.
} 
\label{Fig2}
\end{figure}
%%%%%%%%%%%%%%%%%%%%%%%%%%%%%%%%%%%

\vspace{10mm}

%%%%%  SECTION  4  %%%%%%%%%%%%%%%%%%%%%%%%%%%%%%%%%%%%%%%%%
\section{Summary and discussion}
In the $SU(6) \times SU(2)_R$ string-inspired model with the flavor 
symmetry ${\bf Z}_{19} \times {\bf Z}_{18} \times \tilde{D}_4$, 
we evolve couplings and masses down from the string scale 
$M_S$ using the RG equations. 
In the feasible parameter region of a Yukawa coupling and the 
soft supersymmetry breaking masses, 
the scalar mass squared of the gauge non-singlet matter field 
possibly goes negative slightly below the string scale. 
This implies that the precocious radiative breaking of the gauge 
symmetry $SU(6) \times SU(2)_R$ can occur due to the radiative effect. 
This symmetry breaking triggers off the subsequent symmetry 
breaking as 
\[
  SU(6) \times SU(2)_R \longrightarrow 
   SU(4)_{PS} \times SU(2)_L \times SU(2)_R \longrightarrow 
   G_{SM}.
\]
Thus the present model is in line with the path from the string 
scale physics to the low-energy physics.

In the present model, the large Yukawa coupling which plays 
an important role in the symmetry breaking is identical with 
the colored Higgs coupling. 
This is because the superpotential $W_{Y0}$ of Eq.(\ref{eqn:WY0}) 
induces the colored Higgs mass term 
$z_0 \langle \phi_0 \rangle g_0 g_0^c$ 
below the scale $\langle \phi_0 \rangle$. 
The colored Higgs mass becomes 
$z_0 \langle \phi_0 \rangle = \O(10^{18}{\rm GeV})$. 
This implies that the proton lifetime is more than $10^{36}$yr. 
In contrast to the minimal $SU(5)$ SUGRA GUT model, as to which 
some difficulties have been pointed out concerning 
the proton lifetime,\cite{GN} 
our result is consistent with the present experimental data\cite{SK}. 
The longevity of the proton is in connection with the precocious 
gauge symmetry breaking through the common large Yukawa coupling.

\vspace{10mm}

%%%%%  ACKNOWLEDGEMENTS  %%%%%%%%%%%%%%%%%%%%%%%%%%%%%%%%%%%
\section*{Acknowledgements}
The authors would like to thank Professors C. Hattori, Y. Abe and 
M. Matsunaga for valuable discussions. 
Two of the authors (M. M. and T. M.) are supported in part by 
a Grant-in-Aid for Scientific Research, 
Ministry of Education, Culture, Sports, Science and Technology, 
Japan (No.12047226).

%%%%%  REFERENCES  %%%%%%%%%%%%%%%%%%%%%%%%%%%%%%%%%%%%%%%%%

%%%%  END  %%%%%%%%%%%%%%%%%%%%%%%%%%%%%%%%%%%%%%%%%%%%%%%%%

\end{document}